\documentclass[12pt]{revtex4}

\usepackage{bm}
\usepackage{amsmath}
\usepackage{amssymb}

\newcommand{\etab}{\bm{\eta}}
\newcommand{\bz}{\mathbf{0}}
\newcommand{\bc}{\hat{\mathbf{c}}}
\newcommand{\m}{\mathbf{m}}
\newcommand{\n}{\mathbf{n}}
\newcommand{\petal}{\frac{\partial }{\partial \eta_l}}
\newcommand{\I}{\hat{\mathcal{I}}}
\newcommand{\alphb}{\bm{\alpha}}
\newcommand{\lamdb}{\bm{\lambda}}
\newcommand{\V}{\mathcal{V}}

\def\be{\begin{equation}}
\def\en#1{\label{#1}\end{equation}}

\newcommand{\per}{\mathrm{per}}
\begin{document}

\title[Partial distinguishability and photon counts]{Partial distinguishability and photon counting  probabilities in   linear multiport  devices}

\author{V. S. Shchesnovich}

\address{Centro de Ci\^encias Naturais e Humanas, Universidade Federal do
ABC, Santo Andr\'e,  SP, 09210-170 Brazil }

\begin{abstract} 
Probabilities of photon counts at the output of a multiport optical device are generalised for   optical sources of arbitrary quantum  states  in  partially distinguishable optical modes.  For the single-mode photon sources, the generating function for the probabilities   is a linear combination of the matrix permanents of positive semi-definite Hermitian matrices, where each Hermitian matrix is a Hadamard product of a submatrix of the multiport matrix  and a Hermitian matrix  describing partial distinguishability.  For the multi-mode sources  the generating function is given by  an integral of  the Husimi functions of the sources.  When each photon source outputs exactly a  Fock state,  the obtained expression reduces to the probability formula derived for partially distinguishable photons,   \textit{Physical Review A \textbf{91}, 013844 (2015)}. The derived  probability formula   can be useful in analysing experiments with partially distinguishable sources and error bounds of  experimental Boson Sampling devices.

 \end{abstract}
\maketitle

\vspace{2pc}

\section{Introduction}
\label{sec1}

The purpose of this work is to derive  a formula for photon counts at the output of a  linear unitary optical network with arbitrary photon sources at the input, i.e.,  of arbitrary input quantum states in each port with  arbitrary state of partial distinguishability of the internal optical modes between the ports. Such a result is in order to study   linear quantum optical networks with  important applications in computational complexity   \cite{A1,Valiant}  and the rise of  Boson Sampling with single photons  \cite{AA} with proof-of-principle experiments of several groups \cite{E1,E2,E3,E4,E5,E6,ULO}. The computational complexity of linear  quantum optical networks is not specific to single photons, as it extends to Gaussian quantum states at the input \cite{SCBS,MVS}.  Quantum optical networks for single photons are also at the core of the  universal quantum computing with linear optics \cite{KLM}.

For realistic  photon sources at the input of a  linear multiport,   photon distinguishability   affects ``quantumness'' of such a  device \cite{HOM,Mult1,Ou,Mult2,AHOM}.  For example,   quantum supremacy of the  Boson Sampling  requires  that  indistinguishability of single photons  must be very small  \cite{NDBS,TB}.  In general, the  probability  distribution at the output of  a unitary linear network with   of partial distinguishable single photons (or, more generally,  multi-photon input in Fock states)  is now well studied  \cite{NDBS,PartIndist,Tichy,Rohde,GTh,TL}, nevertheless these results do not extend to  the case of sources producing arbitrary quantum states in the Fock space.   The   formula   for photon counts of an optical field   can be found  in many books (see for instance,  Refs. \cite{MW,VW}). In principle, it can be applied to optical sources producing photons in   partially distinguishable optical modes, which then are sent through a linear network, but such an  application has been not considered in detail.    Due to the recently realised  significance  of linear unitary optical networks for demonstrating quantum supremacy over digital computers, such a result is in order for study of  realistic setups of Boson Sampling  and quantum to classical transition  in linear  optical networks. 
 
The text is organised as follows. In section \ref{sec2} we derive a generating function for photon counting probabilities  at the  output of a multiport device with   arbitrary quantum states in partially distinguishable (internal)  optical modes at the input,  with some details relegated to the appendices.  In section \ref{sec3}, the results are generalised to multi-mode independent sources. In the concluding section \ref{con}, a brief statement of the results is given. 

\section{Generating function for probabilities of photon counts in case of  single-mode optical sources }
\label{sec2}

Consider a unitary linear  $M$-port optical device with arbitrary independent $M$  optical  sources at its input (number of sources can be less than $M$, in this case some of the sources  output the vacuum state, see below). Since the sources are independent, they output quantum states in generally different  set of internal optical modes (which are  the degrees of freedom not affected by the multiport and not resolved by the detectors).  Here we  consider each source to be single-mode. The optical modes of the sources can be accounted for by  introducing a common basis of the optical modes for all $M$ sources, where there are two indices: the  first  index  takes care of the degrees of freedom operated on by a linear multiport (a linear combination of which is resolved by the detectors) and the second for  the degrees of freedom  invariant  under the action of a multiport (the internal modes). The  creation operators of the input, $\hat{a}^\dag_{k,s}$, and  the output, $\hat{b}^\dag_{k,s}$,  basis are related by a multiport optical device described by an unitary matrix $U$ as follows
\be
\hat{b}^\dag_{l,s} = \sum_{l=1}^M U_{k,l}\hat{a}^\dag_{k,s}, \quad l = 1,\ldots,M, 
\en{E1}
where  $s=1,\ldots,M$ enumerates  the basis of the  internal optical modes. We assume that the optical  source  at the input port $k$ of the device outputs an arbitrary single-mode quantum state $\rho^{(k)}$, whose  internal mode is described by the creation operator $\hat{c}^\dag_k$. The latter can be expanded over the input basis
\be
\hat{c}^\dag_k = \sum_{s=1}^M\phi_{k,s}\hat{a}^\dag_{k,s}, \quad \sum_{s=1}^M|\phi_{k,s}|^2 = 1. 
\en{E2}
For the quantum state $\rho^{(k)}$ we thus have 
\be
\rho^{(k)} = \sum_{n,m\ge 0}\ \rho^{(k)}_{n,m}\frac{(\hat{c}^\dag_k)^n|0\rangle\langle0|\hat{c}^m_k}{\sqrt{n!m!}}. 
\en{E3}
Here we do not make assumptions on the  relation between optical modes of different sources (i.e.,   their scalar product  $\sum_{s=1}^M \phi^*_{k,s}\phi_{l,s}$ is arbitrary). 

\textit{For each fixed $k$},  one can complement the  operator $\hat{c}^\dag_k$ to a standard basis of the creation  operators $\hat{c}^\dag_{k,s}$, with $s = 1,\ldots,M$, such that  there is an   unitary transformation between the two sets of  basis operators $\hat{a}^\dag_{k,s}$ and $\hat{c}^\dag_{k,s}$ enumerated by $s=1,\ldots, M$. In other words,  Eq. (\ref{E2}) can be complemented to a unitary transformation for such an  extended set of operators $\hat{c}^\dag_{k,s}$,  $s=1,\ldots,M$. It turns out that  we do  not  need below the exact form  of such an unitary transformation, but only the mere fact that (for each $k$)  the equation  inverse to Eq. (\ref{E2})  can written as follows 
\be
\hat{a}^\dag_{k,s} = \phi^*_{k,s}\hat{c}^\dag_k + \hat{d}^\dag_{k,s},
\en{E4}
where $\hat{d}_{k,s}$ is a linear combination   of creation operators for the internal optical modes orthogonal to that of $\hat{c}_k$. Since the sources produce vacuum in the optical mode described by $\hat{d}_{k,s}$ for $k,s=1,\ldots,M$, there will be no effect of $\hat{d}^\dag_{k,s}$ in Eq. (\ref{E4}) on the photon counts  at the multiport output (thanks to the normal ordering of the creation and annihilation operators  in the photon counting formula, Eqs. (\ref{E5})-(\ref{E6}) below).

The well-known  general formula for photon counts (see, for instance, Refs. \cite{MW,VW}) can be conveniently described by a generating function 
\be
P_\bz(\etab) = \left\langle \mathcal{N}\left\{ \exp\left(- \sum_{k=1}^M \I_k \right)\right\}  \right\rangle,
\en{E5}
where the detector attached at output mode  $k$ is described by the operator $\I_k = \eta_k\sum_{s=1}^M\hat{b}^\dag_{l,s}\hat{b}^{}_{l,s}$,  $0\le \eta_k\le 1$ being its  efficiency (here we take into account that the internal modes are not resolved), $\mathcal{N}$ stands for the normal ordering of creation and annihilation operators,  and $\langle \ldots \rangle$ stands for the averaging with an input quantum state $\rho = \rho^{(1)}\otimes \ldots \otimes \rho^{M}$ of the photon sources. The generating function in Eq. (\ref{E5}) is also the probability of zero photon counts (detecting the vacuum) at the output of a multiport device. The probability of detecting $\m= (m_1,\ldots,m_M)$ photons at the output ports $l=1,\ldots, M$ of an $M$-port is  given  as follows \cite{MW,VW}
\begin{equation}
\label{E6}
P_\m(\etab)=  \left\langle \mathcal{N}\left\{\prod_{l=1}^M\frac{\I_l^{m_l}}{m_l!}  \exp\left(-  \I_l \right)\right\}  \right\rangle =\prod_{l=1}^M\frac{\eta_l^{m_l}}{m_l!} \left(-\petal\right)^{m_l} P_\bz(\etab) .
\end{equation}

First of all, let us express the total detection operator $\sum_{l=1}^m \I_l$ in the $\hat{c}$-mode basis, using Eqs. (\ref{E1}) and (\ref{E4})  we have (dropping on the way the operators $\hat{d}^\dag_{k,s}$ which have no effect on the photon counts)
\begin{eqnarray}
\label{DERIV}
&&\sum_{l=1}^M  \I_l = \sum_{l=1}^M\eta_l \sum_{k,j=1}^M U_{k,l}U^*_{j,l}\sum_{s=1}^M\hat{a}^\dag_{k,s}\hat{a}^{}_{j,s}\nonumber\\
&& = \left[\sum_{k,j=1}^M U_{k,l}\eta_l U^*_{j,l}\right]\sum_{s=1}^M\phi^*_{k,s}\phi_{j,s}\hat{c}^\dag_k\hat{c}^{}_j = \bc^\dag U\Lambda U^\dag \circ \V \bc,
\end{eqnarray}
where    ``$\circ$'' stands for  the by-element (Hadamard) product of two matrices, we have introduced a row-vector of operators $\bc^\dag = (\hat{c}^\dag_1,\ldots,\hat{c}^\dag_M)$, a diagonal matrix $\Lambda = \mathrm{diag}(\eta_1,\ldots,\eta_M)$, and a positive semi-definite Hermitian matrix 
\be
\V_{k,l} \equiv \sum_{s=1}^M \phi^*_{k,s}\phi_{l,s} = \bm{\phi}_k^\dag \bm{\phi}_l^{}
\en{E7}
with   $\bm{\phi}_k \equiv (\phi_{k,1},\ldots,\phi_{k,M})^T$ being the column-vector of the internal optical mode of source $k$ in the common basis $\hat{a}^\dag_{k,s}$.
 
A convenient way to obtain an explicit  expression for the probability in Eq. (\ref{E6}) is to use the Husimi functions and averaging in the coherent basis. For the latter we can  convert Eq. (\ref{E5}) to an equivalent form but involving the anti-normal ordering of boson operators. A general formula for such a conversion (for some  $\bc = (\hat{c}_1,\ldots,\hat{c}_M)^T$) reads (see appendix \ref{appA})
\be
\mathcal{N}\left\{ \exp\left( - \bc^\dag (I_M - H)\bc\right) \right\} = \frac{\mathcal{A}\left\{\exp\left( - \bc^\dag (H^{-1}- I_M)\bc\right) \right\} }{\det(H)},
\en{E8}
where $H$ is an $M$-dimensional positive-definite Hermitian matrix with the eigenvalues bounded by $1$ ($I_M = \mathrm{diag}(1,\ldots,1)$). In our case we have
\be
H = U(I_M-\Lambda)U^\dag \circ \V = I_M - U\Lambda U^\dag \circ \V
\en{H}
 and the generating function becomes 
\begin{eqnarray}
\label{E9}
  P_\bz(\etab)= \left\langle \mathcal{N}\left\{ \exp\left( - \bc^\dag (I_M - H)\bc\right) \right\}  \right\rangle = \left\langle \det(H^{-1})\mathcal{A}\left\{\exp\left( - \bc^\dag (H^{-1}- I_M)\bc\right) \right\}   \right\rangle.
\end{eqnarray}
Introducing a Husimi function for the quantum state of  each optical source, using the coherent state for $\hat{c}_k$-mode,  $\hat{c}_k|\alpha;\bm{\phi_k}\rangle  = \alpha |\alpha;\bm{\phi_k}\rangle$,
\be
Q^{(k)}(\alpha) = \frac{1}{\pi}\langle \alpha;\bm{\phi_k}|\rho^{(k)}|\alpha;\bm{\phi_k}\rangle,
\en{E10}
we obtain the generating function in the form of a multi-mode integral 
\be
P_\bz(\etab)  = \int \left[\prod_{k=1}^M d^2\alpha_k Q^{(k)}(\alpha_k)\right] \frac{\exp\left\{ -\alphb^\dag \left(H^{-1}-I_M\right)\alphb\right\}}{\det(H)},
\en{E11}
where  $\alphb \equiv (\alpha_1, \ldots, \alpha_M)^T$.

In  case of  $N<M$ photon sources (i.e., the rest $M-N$ input ports receive the  optical vacuum), one can easily integrate the vacuum inputs out and reduce the integration in Eq. (\ref{E11})  to  the $N$ non-vacuum sources only. In this case, the matrix  $H$ of  Eq. (\ref{H}) is replaced by a reduced one, built using  $(N\times M)$-submatrix of a multiport matrix $U$ and  the $N$-dimensional submatrix of $\mathcal{V}$  the non-vacuum  sources. Indeed, when some of the sources output the vacuum state,  each such  source has $Q(\alpha) = e^{-|\alpha|^2}/\pi$. As the internal optical mode for the vacuum state can be chosen arbitrarily, we take the   internal modes for  the vacuum Husimi functions to be orthogonal to each other and to the  internal modes of  the rest of the sources. This results in a block-matrix structure for $\V$ and, hence, for  $H$ from Eq.  (\ref{H}):
\be
\V = \left(\begin{array}{cc} \V^{(I)} & 0\\ 0 & \V^{(II)}  \end{array}\right),\quad H = \left(\begin{array}{cc} H^{(I)} & 0\\ 0 & H^{(II)}  \end{array}\right),
\en{E12}
where the superscripts $(I)$ and $(II)$ stand for the non-vacuum and the vacuum sources, respectively,  with the $(II)$-matrices being  diagonal. These properties allow one to  integrate over the $\alpha$-variables corresponding to the vacuum sources (which is a Gaussian integral given in appendix \ref{appB}), where,  taking into account that $\det(H) = \det(H^{(I)}) \det(H^{(II)})$), we obtain the final result \textit{in exactly the same form} as Eq. (\ref{E11}) except that now the integration is only over the $\alpha$-variables of the non-vacuum sources and $H = H^{(I)}$.

Now, given explicit expressions for  Husimi functions of  all optical sources, one can obtain all the needed output probabilities. Below derive other equivalent expressions for the generating function of the output  probabilities,  making connection with the previous results. 

One can integrate  in  the expression  for the generating  function   (\ref{E11}) by performing the  series expansion of the Husimi functions, with the coefficients  being proportional to the terms in the Fock-space expansion of the  quantum state $\rho\equiv \rho^{(1)}\otimes \ldots\otimes  \rho^{(M)}$ of the sources:\begin{equation}
\label{E14}
Q^{(k)}(\alpha)= \frac{1}{\pi}e^{-|\alpha|^2}\sum_{n,m\ge0} \rho^{(k)}_{n,m}\frac{(\alpha^*)^n\alpha^m}{\sqrt{n!m!}}\equiv \frac{1}{\pi}e^{-|\alpha|^2}G^{(k)}(\alpha^*,\alpha).
\end{equation} 
Substituting this expression into  Eq. (\ref{E11})  and observing that the infinite series $G^{(k)}(\alpha_k^*,\alpha_k)$ can be obtained by application of derivatives over some complex dummy variables $\lamdb^\dag = (\lambda^*_1,\ldots,\lambda^*_M)$  in the exponent, we obtain (see appendix \ref{appB})
\begin{eqnarray}
\label{E15}
&&   P_\bz(\etab) =\frac{1}{\det(H)}\int\left[\prod_{k=1}^M \frac{d^2\alpha_k}{\pi} G^{(k)}\left(\frac{\partial}{\partial \lambda_k},\frac{\partial}{\partial \lambda^*_k}\right)\Bigl|_{\lambda_k=0}\right]\exp\left\{ \alphb^\dag H^{-1}\alphb + \alphb^\dag\lamdb + \lamdb^\dag\alphb \right\}\nonumber\\
 && = \left[\prod_{k=1}^M G^{(k)}\left(\frac{\partial}{\partial \lambda_k},\frac{\partial}{\partial \lambda^*_k}\right)\right]\exp\left\{ \lamdb^\dag H \lamdb  \right\}\biggl|_{\lamdb=\bz}.
\end{eqnarray}
Eq. (\ref{E15}) leads to an explicit form of the generating function given  in terms of the matrix permanents of some  Hermitian matrices (with repeated rows and columns, in general) built using rows and columns of  the Hermitian matrix $H$, defined in Eq. (\ref{H}). Indeed, expanding the exponent in Eq. (\ref{E15}) into the Taylor series, we obtain  
\begin{eqnarray}
\label{E16}
&& \prod_{k=1}^M \left(\frac{\partial}{\partial \lambda_k}\right)^{n_k}\left(\frac{\partial}{\partial \lambda^*_k}\right)^{m_k}\biggl|_{\lamdb=\bz}\frac{\left( \lamdb^\dag H \lamdb  \right)^p}{p!}=\delta_{|\m|,p}\delta_{|\n|,p}\sum_{\sigma\in S_p}\prod_{i=1}^pH_{l_i,k_{\sigma(i)}}\nonumber\\
&&=\delta_{|\m|,p}\delta_{|\n|,p} \per(H[\m,\n]),
\end{eqnarray}
where $H[\m,\n]$ is a matrix built from $H$ on rows $l_1,\ldots,l_p$ and columns $k_1,\ldots,k_p$ with repetitions  given by the $\m= (m_1,\ldots,m_M)$ and $\n = (n_1,\ldots,n_M)$, respectively, $S_p$ is the group of permutations of $p$ objects, $|\n| \equiv \sum_{i=1}^M n_i$, and $\per(\ldots)$ is the matrix permanent \cite{Minc}. 
Hence, by the definition of $G^{(k)}$ as an infinite series in Eq. (\ref{E14}), Eq. (\ref{E15}) becomes 
\be
P_\bz(\etab) =\sum_{p\ge0}\sum_{|\n|,|\m|=p}\frac{\per(H[\m,\n])}{\sqrt{\m!\n!}}\prod_{k=1}^M \rho^{(k)}_{n_k,m_k},
\en{E17}
where $\n! \equiv n_1! \ldots  n_M!$.

Eqs. (\ref{E14})-(\ref{E15}) and  (\ref{E17}) constitute  the main result. They can be used for derivation of the specific formulae for photon counting probabilities with general quantum sources and  non-ideal detectors ($\eta_k \ne 1$) by using the prescription in Eq. (\ref{E6}).  Numerical computations of the generating function can turn out to be hard with increase of $N$ and $M$, at least  in some cases, as the example considered below,   due to hardness of the matrix permanent  of positive definite Hermitian matrices \cite{pdH}. The same applies to the formulae for the photon counting probabilities, computational hardness of even approximate calculation of which is at the core of the computational advantage of the Boson Sampling \cite{A1,AA}.

Let us  show, for instance,   that the expression in Eq. (\ref{E17}) reduces to the previously derived probability formula \cite{NDBS,PartIndist} for a fixed number of photons launched in each input port  of a linear multiport, where photons in different input ports  are partially distinguishable. In this case we have 
\be
\rho^{(k)} = \frac{(\hat{c}_k^\dag)^{n_k}|0\rangle\langle0|(\hat{c}_k)^{n_k}}{n_k!}= |n_k;\bm{\phi_k}\rangle\langle n_k;\bm{\phi_k}|,
\en{E18}
i.e., the Fock state in an internal optical mode given by the vector  $\bm{\phi_k}$.  The generating function in this case becomes 
\be 
P_\bz(\etab) = \frac{\per(H[\n,\n])}{\n!}.
\en{E19}
Assuming that we detect all the input photons, $|\n| =N$ (the total number of photons at input),  and using  Eqs. (\ref{E6}), (\ref{H}), and (\ref{E19})  gives 
\begin{eqnarray}
\label{E20}
 &&   P_\m(\etab) = \frac{1}{\n!}\prod_{l=1}^M\frac{\eta_l^{m_l}}{m_l!} \left(-\petal\right)^{m_l} \per(H[\n,\n])= \frac{1}{\m!\n!}\prod_{l=1}^M\eta_l^{m_l} \left(-\petal\right)^{m_l} \sum_{\sigma\in S_N}\prod_{i=1}^N H_{k_i,k_{\sigma(i)}}\nonumber\\
&&   = \frac{1}{\m!\n!}\sum_{\sigma,\tau\in S_N} \prod_{i=1}^N\eta_{l_i} U_{k_i,l_{\tau(i)}}U^*_{k_{\sigma(i)},l_{\tau(i)}}\V_{k_i,k_{\sigma(i)}}=\frac{\etab^{\m}}{\m!\n!}\sum_{\sigma_{1,2}\in S_N}J(\sigma_2\sigma^{-1}_1)\prod_{i=1}^NU_{k_{\sigma_1(i)},l_i}U^*_{k_{\sigma_2(i)},l_i},\nonumber\\
\end{eqnarray}
where $k_1,\ldots,k_N$ and $l_1,\ldots,l_N$ are the input and output ports of a multiport (generally, with repetitions) corresponding to the occupations $\n$ and $\m$, respectively, $\etab^\m = \eta_1^{m_1} \ldots \eta_M^{m_M}$, $\sigma_1 = \tau^{-1}$, $\sigma_2 = \sigma\tau^{-1}$, and 
\be
J(\sigma) \equiv \prod_{i=1}^N \V_{k_i,k_{\sigma(i)}} = \prod_{i=1}^N\bm{\phi}_{k_i}^\dag \bm{\phi}_{k_{\sigma(i)}}^{} = \bm{\Phi}^\dag P_{\sigma^{-1}}\bm{\Phi}
\en{E21}
where we have introduced the tensor-product  of the  internal  states of photons  
\be
\bm{\Phi} \equiv \bm{\phi}_{k_1}\otimes\ldots \otimes \bm{\phi}_{k_N} =  \left(\bm{\phi}_1\right)^{\otimes n_1}\otimes \ldots \otimes \left(\bm{\phi}_M\right)^{\otimes n_M}.  
\en{E22}
 and the operator representation $P_\sigma$ of a permutation $\sigma$. 
The result in Eqs. (\ref{E20}) and (\ref{E21}) reproduces that of Ref. \cite{PartIndist} when $\eta_k = 1$ (to make a comparison  clearer, note that  $U$ here is equivalent to  $U^\dag$ in Ref. \cite{PartIndist}, and  that $\bm{\Phi}$ is the vector of expansion coefficients in a basis of internal states of photons, therefore $P_{\sigma^{-1}}$ in Eq. (\ref{E21}) corresponds to $P_\sigma$ in the definition of $J(\sigma)$ of Ref. \cite{PartIndist}).

\section{Multi-mode independent sources} 
\label{sec3}

When photon sources output  multi-mode quantum states, one has to employ a basis for the internal modes.  We can choose any basis, since, in general, no preferred basis exists (only in the case of  single-mode sources, one can select a special basis for each source, as in  Eq. (\ref{E3})). 
Therefore, we assume that there is a common basis, $\hat{a}^\dag_{k,s}$, of the internal modes and that it is of a finite dimension $s=1,\ldots,d$. Eq. (\ref{E3}) is replaced in this case by 
\be
\rho^{(k)} = \sum_{\n,\m} \rho^{(k)}_{\n,\m} \frac{\prod_{s=1}^d(\hat{a}^\dag_{k,s})^{n_s}|0\rangle\langle0|\prod_{s=1}^d\hat{a}^{m_s}_{k,s}}{\sqrt{\n!\m!}},
\en{E23}
where $\n = (n_1,\ldots,n_d)$ and $\n! = n_1!\cdot \ldots\cdot n_d!$. One can proceed now in a manner similar to that of section \ref{sec2}, with the two differences. First, source  $k$ is now described by a generalised  Husimi function $Q^{(k)}(\alphb^{(k)})$ of $d$ complex variables $\alphb^{(k)} \equiv (\alpha_1, \ldots,\alpha_d)^T$, where
\be
Q^{(k)} = \frac{1}{\pi^d} \langle \alphb^{(k)}|\rho^{(k)}|\alphb^{(k)}\rangle, \quad |\alphb^{(k)}\rangle \equiv |\alpha^{(k)}_1\rangle\otimes \ldots \otimes |\alpha^{(k)}_d\rangle,
\en{E24}
where $\hat{a}_{k,s}|\alpha^{(k)}_s\rangle = \alpha^{(k)}_s|\alpha^{(k)}_s\rangle$. Second, since we use a common basis, the $M$-dimensional Hermitian matrix $H$ of Eq. (\ref{H}) is replaced by the $M\times d$-dimensional one (note the tensor product below, and not the Hadamard product  as in Eq. (\ref{E3}))
\be
\mathcal{H} = U (I_M - \Lambda) U^\dag \otimes I_d.
\en{E25}
The rest of the derivation is just a mere repetition of the steps made in section \ref{sec2}. We obtain the following result for the generating function of the output probability distribution
\be
P_{\mathbf{0}}(\etab)  = \frac{1}{\prod_{k=1}^M(1-\eta_k)^d}\int \left[\prod_{k=1}^M d^2\alphb^{(k)} Q^{(k)}(\alphb^{(k)}) \right]\exp\left\{ -\alphb^\dag \left(\mathcal{H}^{-1}-I_{M\times d}\right)\alphb\right\}
\en{E26} 
where $\alphb^T \equiv (\alphb^{(1)},\ldots,\alphb^{(M)})^T$ and we have used that $\mathrm{det}(\mathcal{H}) = \prod_{k=1}^M(1-\eta_k)^d$.  
Similarly as for the single-mode case in section \ref{sec2}, for $N<M$ sources  (with  $M-N$ input ports receiving the optical vacuum), one can easily integrate the vacuum inputs out and reduce the integration in Eq. (\ref{E26})  to $N$ of  $\alphb^{(k)}$ corresponding to non-vacuum sources  only, in quite a similar fashion.

Note that the internal-mode configuration of the sources is now contained in the Husimi functions themselves and not in the Hermitian matrix $\mathcal{H}$,  in contrast to the single-mode case, Eq.~(\ref{E11}).  Therefore, though one can integrate in Eq. (\ref{E26}),  similar as it was done in Eqs. (\ref{E14})-(\ref{E17}) of section \ref{sec2}, the expression will be quite cumbersome, in general. This is the main difference between the single-mode and the multi-mode sources. The reason for this is the non-existence of a preferred basis: since, in general, any $M\times d$-dimensional basis of boson operators  can be used in the multi-mode case, there is no point in introducing  the operators $\hat{c}_{k,s}$ (as a generalisation of $\hat{c}_k$ of Eq. (\ref{E2})) specific for each source in the multi-mode case.

Finally,  recalling  the example of single-photon sources, analysed in section \ref{sec2}, there is but a marginal  additional generality coming from Eq. (\ref{E26}) in derivation of the previous result of Ref. \cite{PartIndist}. Indeed, since the number of photons per input is fixed, one can  always expand  the corresponding density matrix $\rho^{(k)}$ of Eq. (\ref{E23}) as a positive combination, where each term is a product of single-mode ones of Eq. (\ref{E3}) (with different internal modes),  and use  the approach of section \ref{sec2}. In this case, Eq. (\ref{E21}) is replaced by the most general one involving the ``internal density matrix" instead of the scalar product of the internal modes (see, for details, Ref. \cite{PartIndist}).

\section{Conclusion}
\label{con}
The main result of this work is a  generating function for the  photon counts at the output of a linear optical multiport with arbitrary photon sources at the input. For the single-mode  optical sources, the latter is  a linear combination of matrix permanents of positive semi-definite Hermitian matrices, where each Hermitian matrix is  given as the Hadamard product of a submatrix of  the unitary matrix of a multiport device and a Hermitian matrix  describing partial distinguishability of the  internal optical modes of photon   sources.  For the case of multi-mode sources we have, in general, the generating function as an integral of the (multidimensional) Husimi functions of the sources and a Gaussian exponent describing the action of a linear multiport and photon   detection stage  (without resolving the internal optical modes).  When each photon source outputs exactly a  Fock state,  the obtained expression reduces to the probability formula derived before for partially distinguishable  photons. The results    can be useful in analysing the interference  experiments with general  optical sources of photons  with controlled distinguishability  and for derivation of error bounds for the  experimental Boson Sampling devices.

\section{Acknowledgements}

The research was supported by the National Council for Scientific and Technological Development (CNPq) of Brazil,  grant  304129/2015-1, and by  the S{\~a}o Paulo Research Foundation   (FAPESP), grant 2015/23296-8. 
 
\appendix
\section{Relation between the  normal and the anti-normal ordering of an exponent of a quadratic form in boson  operators}
\label{appA}

Let us first  show the following relation
\be
\mathcal{N}\left\{ \exp\left( -\xi \hat{b}^\dag \hat{b} \right) \right\} = \frac{\mathcal{A}\left\{ \exp\left(- \lambda \hat{b}^\dag \hat{b} \right) \right\} }{1-\xi},
\en{A1}
where $0< \xi< 1$ and $\lambda = \xi/(1-\xi)$.  To this end we observe that for two Fock states $|n\rangle $ and $|m\rangle$
\begin{eqnarray}
\label{A2}
&&\langle m|\mathcal{N}\left\{ \exp\left( -\xi \hat{b}^\dag \hat{b} \right) \right\}|n\rangle = \sum_{k=0}^\infty \frac{(-\xi)^k}{k!}\langle m|(\hat{b}^\dag)^kb^k|n\rangle \nonumber\\
&& = \delta_{m,n} \sum_{k=0}^n\frac{(-\xi)^k n!}{(n-k)!k!} = \delta_{m,n}(1-\xi)^n,
\end{eqnarray}
similarly 
\begin{eqnarray}
\label{A3}
&&\langle m|\mathcal{A}\left\{ \exp\left( -\lambda \hat{b}^\dag \hat{b} \right) \right\}|n\rangle = \sum_{k=0}^\infty \frac{(-\lambda)^k}{k!}\langle m|\hat{b}^k(\hat{b}^\dag)^k|n\rangle \nonumber\\
&& = \delta_{m,n} \sum_{k=0}^n\frac{(-\lambda)^k (k+n)!}{n!k!} = \delta_{m,n}(1+\lambda)^{-n-1},
\end{eqnarray}
where we have used that 
\be
\sum_{k=0}^\infty \frac{\lambda^k(k+n)!}{n!k!} = (1+\lambda)^{-n-1}.
\en{A4}
Eqs. (\ref{A2}) and (\ref{A3}) immediately give  Eq. (\ref{A1}). 

Consider now Eq. (\ref{E8}) of section \ref{sec2}. Let $H^\dag = H$ and $0< H < I$. We can diagonalize the positive-definite Hermitian matrix $H = V^\dag D V$, where $D = \mathrm{diag}(1-\xi_1,\ldots, 1-\xi_M)$, $0<\xi_k<1$,  and $V^\dag V=I$. Using $V$ for a canonical transformation of the operators $\hat{c}_k$  to a new basis $\hat{b}_k$, $k=1,\ldots,M$,
\be
\hat{b}_k = \sum_{l=1}^M V_{k,l} \hat{c}_l, 
\en{A5}
and observing that $\det(H) = \prod_{k=1}^M(1-\xi_k)$,  we reduce Eq. (\ref{E8}) to an operator  product form, where each element in the product is a relation given by Eq. (\ref{A1}). This finishes the proof of Eq. (\ref{E8}).

\section{A Gaussian integral}
\label{appB}
Here we give for a reference a Gaussian integral which is used in the main text
\be
\int\prod_{k=1}^M \frac{d^2\alpha_k}{\pi} \exp\left( - \alphb^\dag A\alphb + \lamdb^\dag \alphb + \alphb^\dag \bm{\mu}\right) = \frac{\exp\left(\lamdb^\dag A^{-1}\bm{\mu}\right)}{\det(A)}
\en{B1}
where $A$ is a positive definite  Hermitian matrix.  Eq. (\ref{B1}) is derived by first diagonalizing the matrix $A = \mathcal{U}^\dag \mathrm{diag}(a_1,\ldots,a_M) \mathcal{U}$, making a change of integration variable $\mathbf{z} = \mathcal{U}\alphb$, and invoking the following result 
\be
\int \frac{d^2 z}{\pi} \exp\left(-a|z|^2 +\lambda^* z + z^*\mu \right) = \frac{1}{a}\exp\left( \frac{\lambda^*\mu}{a}\right).
\en{B2}

\end{document}